\documentstyle[10pt]{article}

\begin{document}
\centerline{\Large\bf Teardrop and heart orbits}
\vspace{0.1cm}
\centerline{\Large\bf of a swinging Atwood's Machine}
\vspace{0.2cm}
\centerline{Nicholas B. Tufillaro}
\vspace{0.2cm}
\centerline{{\em Center for Nonlinear Studies and T13, MS-B258}}
\vspace{0.1cm}
\centerline{{\em Los Alamos National Laboratory, Los Alamos, New Mexico 87545 USA}}

\bigskip\bigskip

\noindent{\bf Abstract:}
An exact solution is presented for
a swinging Atwood's machine. This teardrop-heart 
orbit is constructed using 
Hamilton-Jacobi theory. The example nicely illustrates the utility
of the Hamilton-Jacobi method 
for finding solutions to nonlinear mechanical systems 
when more elementary techniques fail.
\par
\vspace{2cm}
Date: 14 February 1993\\
\\
\\
To be submitted to: {\ \ \it The American Journal of Physics.}

\vspace{1cm}
\noindent

\newpage

A swinging Atwood's machine (SAM) is an
ordinary Atwood's machine in which, however, 
one of the weights can swing in a plane [1].
In the polar coordinates $(r , \theta)$, 
the nonlinear Hamiltonian describing the dynamics
of SAM is
\begin{equation}
H = T + V = {1 \over 2} (m+M) \dot  r^2 + {1 \over 2} m r^2 \dot\theta^2
+ g r (M + m \cos  \theta  ), \label{(1)}
\end{equation}
where 
$M$ is the mass of the nonswinging bob,
$m$ is the mass of the
swinging bob, 
$g$ is the gravitational acceleration,
and the polar angle $\theta$ is measured from the negative 
vertical y-axis as shown in Figure 1(a).
When the mass ratio $\mu = M/m = 3$, 
numerical studies of SAM reveal an unexpected feature.
To wit, all orbits which start at the origin (with
initial conditions 
$r = \epsilon,\ \dot r = v,\ \theta = \theta_0,\ \dot \theta = 0$)
{\it execute one symmetrical loop and return to the origin, 
no matter what the launch angle ($\theta_0$) or speed ($v$).}
When the launch angle is small, the orbit is called a $teardrop$,
and when the launch angle is 
larger (say $\theta_0 > \pi/2)$ the orbit is called a $heart$ (see Figure 1) [2].  

Since the SAM Hamiltonian is nonlinear, 
we generally expect that it will be
impossible to find (simple) exact solutions for SAM [3]. 
In this paper I
show that the ``teardrop-heart'' 
orbit of SAM is an exception to this rule.
Specifically, I use Hamilton-Jacobi theory
to construct an exact solution for the
teardrop-heart orbit. This example would make
a nice addition to an advanced mechanics class
since it illustrates the power of the 
Hamilton-Jacobi method to arrive at an exact solution
which apparently alludes more elementary methods.

Recall that the game of Hamilton-Jacobi theory 
is to find some canonical transformation which
will separate variables in the Hamilton-Jacobi 
equation. A fortuitous choice for the
SAM Hamiltonian is the point transformation
\begin{equation}
r = {1 \over 2} (\xi^2 + \eta^2),
\end{equation}
\begin{equation}
\theta = 2 \arctan [ (\xi^2 - \eta^2) / 2 \xi \eta ].
\end{equation}
This transformation compresses the angular variable by a factor of two, and then
changes from polar coordinates to parabolic coordinates with parabolas centered about the
horizontal $x$-axis [4]. The inverse transformation is
\begin{equation}
\xi = \sqrt r \sqrt{ 1 +\sin(\theta/ 2)},
\end{equation}
\begin{equation}
\eta = \pm \sqrt r \sqrt{1 - \sin(\theta/2)}.
\end{equation}
If I further set $M = 3m$ (and $M = 1$) as is done throughout
the rest of this paper I find that the generalized momentum
in the new coordinate system is
\begin{equation}  
p_\xi = 4  \dot \xi (\xi^2 + \eta^2) =
{\partial S_\xi \over \partial \xi},
\ \ \ p_\eta = 4 \dot \eta (\xi^2 + \eta^2) =
{\partial S_\eta \over \partial \eta},
\end{equation}
from which the (time-dependent) Hamiltion-Jacobi equation is calculated as
\begin{equation}
{\partial {\cal S} \over {\partial t}} +
{1 \over (\xi^2 + \eta^2)}
\left \{
{1 \over 8} 
\left [ 
\left({\partial {\cal S} \over {\partial \xi}}\right)^2 +
\left({\partial {\cal S} \over {\partial \eta}}\right)^2 
\right ] 
+
2 g ( \xi^4 + \eta^4)  
\right \}
= 0 .
\end{equation}

The equation for the generating function is now separable 
and can be expressed in the form
\begin{equation}
{\cal S}(\xi, \eta, t) = S_\xi (\xi) + S_\eta (\eta) + S(t).
\end{equation}
Separating the time and space parts of Eq.\ (7) I find that the
first separation constant is just the total energy, $E$:
\begin{equation}
{{\partial \cal S} \over {\partial t}} = -E,
\end{equation}
\begin{equation}
{1 \over (\xi^2 + \eta^2)}
\left \{
{1 \over 8} 
{\left [ 
\left({\partial {\cal S} \over {\partial \xi}}\right)^2 +
\left({\partial {\cal S} \over {\partial \eta}}\right)^2 
\right ]} 
+
2 g ( \xi^4 + \eta^4) 
\right \} 
 = E .
\end{equation}
Integration of Eq.\ (9) yields
\begin{equation}
{\cal S}(t) = -E t .
\end{equation}
Equation (10) can be written as
\begin{equation}
{1 \over 8} 
{\left [ 
\left({\partial {\cal S} \over {\partial \xi}}\right)^2 +
\left({\partial {\cal S} \over {\partial \eta}}\right)^2 
\right ] 
+
2 g ( \xi^4 + \eta^4) 
}  = E(\xi^2 + \eta^2).
\end{equation} 
Next I can separate the spatial part of the Hamilton-Jacobi (Eq.\ (12))
equation giving
\begin{equation}
\left ( {\partial \cal S \over \partial \xi}\right)^2 
+ 16 g \xi^4 - 8 E \xi^2 
= I,
\end{equation}
\begin{equation}
- \left ( {\partial \cal S \over \partial \eta}\right)^2 
- 16 g \eta^4 + 8 E \eta^2 
= I,
\end{equation}
where I is the spatial separation constant.
The complete solution to the
Hamilton-Jacobi equation is then given by
\begin{equation}
{\cal S} =
\int (I +  8E\xi^2 - 16 g \xi^4)^{1\over 2} d\xi + 
\int (-I + 8E\eta^2 -16g\eta^4)^{1\over 2} d\eta - Et.
\end{equation}

To ``solve'' the original problem I do not need to
integrate Eq.\ (15) directly, but rather the {\it orbit equation}:
\begin{equation}
{{\partial \cal S} \over {\partial I}} =
{1 \over 2}
\int {{1} \over {(I + 8E\xi^2 -16g\xi^4)^{1\over2}}} d \xi 
-
{1 \over 2}
\int {{1} \over {(-I + 8E\eta^2 -16g\eta^4)^{1\over2}}} d \eta
= \alpha
\end{equation}
which provides an implicit relation between $\xi$ and $\eta$ (and, therefore,
$r$ and $\theta$), and the parametrized {\it trajectory equation}:
\begin{equation}
{{\partial \cal S} \over {\partial E}} =
\int {{4 \xi^2} \over {(I + 8E\xi^2 -16g\xi^4)^{1\over2}}} d \xi 
+
\int {{4 \eta^2} \over {(-I + 8E\eta^2 -16g\eta^4)^{1\over2}}} d \eta
= t + \beta 
\end{equation}
where $\alpha$ and $\beta$ are constants of integration which may, in general,
be a complicated mix of the original initial conditions.  

Just as important, I can also solve for the new constant of
the motion, $I$, by adding Eqs.\ (13) and (14), 
and then eliminating the energy constant to get
\begin{equation}
I = 2 (\xi^2 + \eta^2)
(\eta^2 \dot \xi^2 - \xi^2 \dot \eta^2)
+
2 g \xi^2 \eta^2 [(\xi^2 -\eta^2) / (\xi^2 + \eta^2)].
\end{equation}
In the original polar coordinate system 
(apply Eqs.\ (4) and (5) to Eq.\ (18)) the new
constant of motion reads
\begin{equation}
I(r, \dot r, \theta, \dot \theta) =
{r^2 \dot \theta}
{
\left [
\dot r \cos \left ( \theta \over 2 \right) -
{{r \dot \theta} \over 2} \sin \left ( \theta \over 2 \right )
\right ]
}
+
g r^2 \sin \left ( \theta \over 2 \right )
\cos^2 \left ( \theta \over 2 \right ) .
\end{equation}

In principal the orbit and trajectory equations arising from the
generating function represent a complete solution to the original
problem --- 
in this case finding the motion of SAM when $\mu = 3$. 
In practice, though, there
are still two difficult obstacles to overcome. First, actually finding
explicit integrals to
the orbit and trajectory equations may not be possible, and second, the
resulting relations are often, at best, only implicit functions
which can not be inverted for the relevant variables.

In this particular example, the first of these obstacles can be
overcome --- the orbit and trajectory equations can be explicitly integrated
with Elliptic functions. However, instead of using special functions
to arrive at a solution, I would next like to turn my attention
to a specific solution --- the teardrop-heart orbits --- and show
that for this special class of orbits it is possible to obtain
a simple and elementary solution for the motion of SAM.
 
The key observation used in arriving at this elementary solution is as
follows: 
{\it for all teardrop-heart 
orbits the constant of motion ``$I$'' is equal to zero}.
To see why this last observation is true consider the teardrop-heart orbit
as it crosses the negative vertical axis (see Figure 1). 
At the axis $\theta = 0$ and $\dot r = 0$. Plugging these values into
Eq.\ (19) it follows that $I(r, \dot r, \theta, \dot \theta) = 0$ for all $t$.   
Next I show that when $I = 0$ the integrals in the orbit and trajectory
equations can be solved by elementary functions, 
and the resulting functions
can be inverted to arrive at a 
simple orbit equation relating $r$ and $\theta$.

Examining the special case $I = 0$, the orbit equation (Eq.\ (16)) becomes
\begin{equation}
2 \alpha =   
\int {{1} \over {(8E\xi^2 -16g\xi^4)^{1\over2}}} d \xi 
-
{1 \over 2}
\int {{1} \over {(8E\eta^2 -16g\eta^4)^{1\over2}}} d \eta
\end{equation}
which can be integrated straight away  to give [5]:
\begin{equation}
2 \alpha =
{{1}\over{\sqrt{8E}}}
\left \{
\left [
\ln \left (
\sqrt{2E - 4g\xi^2 }
-\sqrt{2E}
\right)
- \ln(\xi)
\right ]
- \left [
\ln
\left(
\sqrt{2E - 4g\eta^2}
-\sqrt{2 E}
\right)
-
\ln(\eta)
\right ]
\right \}.
\end{equation}

After some additional algebraic manipulations I find the
implicit orbit equation in $\xi$ and $\eta$ to be
\begin{equation}
a = {{\eta}\over{\xi}}
{{\sqrt{1-k\xi^2} - 1}\over{\sqrt{1-k\eta^2} - 1}},
\end{equation}
where the positive constants $a$ and $k$ are defined by
\begin{equation} 
a = e^{2\alpha\sqrt{8E}},\ \ \ 
k = {{2g}\over{E}}.
\end{equation}
Using the inverse point transformation 
Eqs.\ (4) and (5) applied to Eq.\ (22) I get
the implicit orbit equation in polar coordinates as
\begin{equation}
\pm a = 
{
{\sqrt{1 - \sin {\theta\over2} }}  
\over 
{\sqrt{1 + \sin {\theta\over2} }}
}
{
{\sqrt{1-kr(1+\sin\theta)}-1}
\over
{\sqrt{1-kr(1-\sin\theta)}-1}
}.
\end{equation}

At this point it is helpful to pause and check this result by considering
the limiting case of a teardrop orbit. 
Namely, the solution of Eq.\ (24) in the limit
of small $\theta$. Expanding Eq.\ (24) in $\theta$ about zero I find [6]:
\begin{equation}
\pm a = 1 + 
{
{\theta}
\over
{2 \sqrt{1-kr}}
}
+ {\cal O}(\theta^2).
\end{equation}  
Or, written another way 
\begin{equation}
r(\theta) \approx  r_0 \left [
1 - \left ( {{\theta}\over{\theta_0}} \right )^2 
\right ],
\ \ \ r_0 = 1/k, 
\ \theta_0 = 2(-1 \pm a),
\end{equation}
which indeed produces the teardrop 
shaped orbit shown in Figure 1(a).

Now I return to the main stream of my presentation by attempting
to find an explicit orbit equation. Rewrite Eq.\ (22) in the form
$a\xi\sqrt{1-k\eta^2} - \eta \sqrt{1-k\xi^2} = a\xi - \eta$, 
square both sides, and collect like terms to discover 
$2a - k\xi\eta - ka^2\xi\eta = 2a \sqrt{1 -k\xi^2 - k\eta^2 + k^2\xi^2\eta^2}$.
Next notice (from Eqs.\ (4) and (5)) that 
$\xi\eta = r\cos\theta$.
Evidently, 
$
(k^2 + k^2a^4 - 2a^2k^2) r^2 \cos^2\left({\theta\over2}\right)+
(-4ak-4a^3k)r\cos\left({\theta\over2}\right)
+
(4 a^2 k)r \left [1+\sin\left({\theta\over2}\right) \right ] 
+ 
(4 a^2 k) r\left [1-\sin \left({\theta\over2}\right)\right ] = 0.  
$
Additional algebraic simplification 
results in the exact explicit {\it teardrop-heart orbit equation }
\begin{equation}
r(\theta) = 
{{4a}\over{k}}  \cdot  
{
   { 
     (a^2+1)
     \left\vert\cos\left({\theta\over2}\right)\right\vert - 2a
   }
\over
   {
     (a^2-1)^2 \cos^2\left({\theta\over2}\right)
   }
},
\ \ \ -\theta_0 < \theta < \theta_0,
\end{equation}
that is, the teardrop-heart orbit occurs for positive values of $r$, and this 
condition places the above restriction on $\theta$ where (at $r = 0$)
\begin{equation}
\theta_0 = 2 \arccos\left({{2a}\over{a^2+1}}\right).
\end{equation}
Also, when the orbit crosses the negative vertical axis 
its length is (at $\theta=0$)
\begin{equation}
r_0 = {{4a}\over{k}} \cdot
{{1}\over{(a+1)^2}} .
\end{equation}
That's it.  

\vspace{1cm}

\noindent
Acknowledgements: 
It is a pleasure to thank 
Don Stark and Nicholas Robidoux for their algebraic insights.
\newpage
\centerline{\bf References}
\vspace{1ex}\vspace{1ex}
\noindent
[1].  David J. Griffiths and Tyler A. Abbott, Comment on ``A surprising
mechanics demonstration,'' Am. J. Phys. {\bf 60} (10), 951-953 (1992). 
\vspace{1ex}

\noindent 
[2]. Nicholas B. Tufillaro, {\it Smiles and Teardrops},
B.\ A.\ Thesis, Reed College, Portland, Oregon 97202 (1982). 
\vspace{1ex}

\noindent
[3]. Nicholas B. Tufillaro, Tyler Abbott, and Jeremiah Reilly, 
{\it An experimental approach
to nonlinear dynamics and chaos} (Addison-Wesley, Reading, MA, 1992).
\noindent\vspace{1ex}                                                                       

\noindent
[4]. The identity $\cos(2\theta) = (1-\tan^2\theta)/(1+\tan^2\theta)$ is useful
in making this change of coordinates.
In Cartesian coordinates the transformation is 
$\xi = [\sqrt{x^2 + y^2} + x]^{1\over 2}$ and 
$\eta =  \pm[\sqrt{x^2 + y^2} - x]^{1 \over 2}$.
\noindent\vspace{1ex}

\noindent
[5]. These integrals are quickly done with a symbolic program such as
{\it Mathematica} or {\it Axiom}. The latter program gives:
$$
\int {{1} \over {(8 E \xi^2 - 16 g \xi^4)^{1\over2}}} = 
{{1}\over{\sqrt {8E}}} \left [
\log \left ( \sqrt {2E - 4g\xi^2} - \sqrt{2E} \right ) - \log{(\xi)} \right ],
$$
and
$$
\int {{8\xi^2} \over {(8 E \xi^2 - 16 g \xi^4)^{1\over2}}} =
-{{\sqrt 2}\over {g}} \left [ \sqrt {E - 2 g \xi^2} \right ].
$$

\vspace{1ex}
\noindent
[6]. This expansion is easily done with a symbolic mathematics 
program. In {\it Mathematica} the command 
``Simplify[Series[(Eq.\ (24)), \{$\theta$, 0, 2\}]''
results in
$$
1 + {\theta\over {2\,{\sqrt{1 - k\,r}}}} + 
  {{\left( -2 + 2\,k\,r - 2\,{\sqrt{1 - k\,r}} + 
        k\,r\,{\sqrt{1 - k\,r}} \right) \,{\theta^2}}\over 
    {8\,{\sqrt{1 - k\,r}}\,\left( -1 + k\,r \right) \,
      {{\left( 1 + {\sqrt{1 - k\,r}} \right) }^2}}} + {{{\rm\cal O}(\theta^3)}}.
$$ 

\vspace{1ex}
\noindent

\newpage
\centerline{\bf Figure Captions}
\vspace{1ex}
\noindent
{\bf Figure 1:} 
(a) Teardrop, (b), (c) heart and 
(d) a rabbit ear orbit
of a swinging Atwood's machine (SAM) when $\mu = M/m = 3$.
\par
\vspace{1ex}
\noindent

\end{document}